\documentclass[10pt,conference]{IEEEtran}
\IEEEoverridecommandlockouts

\usepackage{multirow,multicol}

\usepackage[flushleft]{threeparttable}

\usepackage[linesnumbered,titlenumbered,ruled,vlined]{algorithm2e}

\usepackage{tikz} 
\usepackage{amsmath}
\usepackage{hyperref}
\usepackage[capitalize]{cleveref} 
\usepackage{xspace}
\usepackage{fancyvrb} 
\usepackage[T1]{fontenc}
\usepackage{enumitem}
\usepackage{txfonts}
\usepackage{nicefrac}
\usepackage[svgnames]{xcolor}
\usepackage{siunitx}
\usepackage{array,framed}
\usepackage{booktabs}
\usepackage{csquotes}
\usepackage{
  float,
  epsfig,
  wrapfig,
  graphicx,
  subcaption
}
\usepackage{textcomp,amssymb}
\usepackage{setspace}
\usepackage{latexsym,fancyhdr,url}
\usepackage[framemethod=default]{mdframed}
\usepackage{wasysym}

\usepackage{xparse} 
\usepackage{csvsimple}
\usepackage{rotating}

\usepackage{balance}

\usepackage{fontawesome}
\usepackage{comment}
\usepackage{svg}
\newcommand{\sys}{\mbox{\texttt{SafePlanner}}\xspace}

\newcommand{\nibf}{\noindent\textbf}
\newcommand{\etal}{\textit{et al}.\xspace}

\SetCommentSty{mycommfont}

\mdfdefinestyle{answerbox}{align=center,innerleftmargin=.15cm,linecolor=black,%
                    linewidth=1pt,backgroundcolor=Gainsboro}
\newmdenv[style=answerbox]{answerbox}

\SetKwInOut{Parameter}{Parameters}

\begin{document}

\title{SafePlanner: Testing Safety of the Automated Driving System Plan Model}

\author{
\IEEEauthorblockN{
Dohyun Kim\textsuperscript{1},
Sanggu Han\textsuperscript{1},
Sangmin Woo\textsuperscript{1},
Joonha Jang\textsuperscript{2},
Jaehoon Kim\textsuperscript{1},
Changhun Song\textsuperscript{1},
Yongdae Kim\textsuperscript{1}
}
\IEEEauthorblockA{
\textsuperscript{1}KAIST\quad
\textsuperscript{2}Korea Air Force Academy\\
\{dohyunjk, 139, wsm723, jaehoon.kim99, songch, yongdaek\}@kaist.ac.kr\quad
cyber040946@gmail.com
}
}

\maketitle

\begin{abstract}

In this work, we present \sys, a systematic testing framework for identifying safety-critical flaws in the Plan model of Automated Driving Systems (ADS).
\sys targets two core challenges: generating structurally meaningful test scenarios and detecting hazardous planning behaviors.
To maximize coverage, \sys performs a structural analysis of the Plan model implementation—specifically, its scene-transition logic and hierarchical control flow—and uses this insight to extract feasible scene transitions from code.
It then composes test scenarios by combining these transitions with non-player vehicle (NPC) behaviors.
Guided fuzzing is applied to explore the behavioral space of the Plan model under these scenarios. We evaluate \sys on Baidu Apollo, a production-grade level 4 ADS.
It generates 20,635 test cases and detects 520 hazardous behaviors, grouped into 15 root causes through manual analysis.
For four of these, we applied patches based on our analysis; the issues disappeared, and no apparent side effects were observed.
\sys achieves 83.63\% function and 63.22\% decision coverage on the Plan model, outperforming baselines in both bug discovery and efficiency.
\end{abstract}


\section{Introduction}
\label{sec:intro}

Automated Driving Systems (ADSes) are being rapidly deployed to improve road safety and mobility.
However, recent failures of ADS-equipped vehicles have highlighted the risks of incomplete or erroneous decision logic.
According to the U.S. National Highway Traffic Safety Administration (NHTSA), 392 crashes involving ADS-equipped vehicles were reported between July 2021 and May 2022~\cite{nhtsa_report}, underscoring the urgency of comprehensive and systematic testing frameworks to ensure safety and reliability.

A core component of an ADS is the Plan model, which determines the vehicle’s motion by generating trajectories based on perceived environment and navigation intent.
Planning failures can directly lead to unsafe behaviors, such as collisions, improper lane changes, or failure to yield. Such failures often arise from incorrect assumptions, inadequate handling of corner cases, or flawed logic propagation in dynamically changing scenes.
The planning task is complicated by the fact that multiple reasonable behaviors may exist for a given driving situation.
For example, when encountering a stationary vehicle ahead, an ADS might wait, perform a lane change, or temporarily merge and return.
Each behavior may be valid depending on context, but incomplete or biased logic can still result in safety-critical outcomes.

This behavioral ambiguity, coupled with the lack of implementation constraints in ADS standards, shifts design decisions to developers, whose heuristics or simplifications may leave the system vulnerable.
Plan model malfunctions can manifest as either aggressive behaviors (e.g., overtaking in unsafe gaps) or overly conservative ones (e.g., stopping indefinitely), both of which threaten usability and safety.
As Plan models are tightly coupled with vehicle control, their logic must be tested rigorously.

\begin{figure}[t]
    \centerline{\includegraphics[width=\columnwidth]{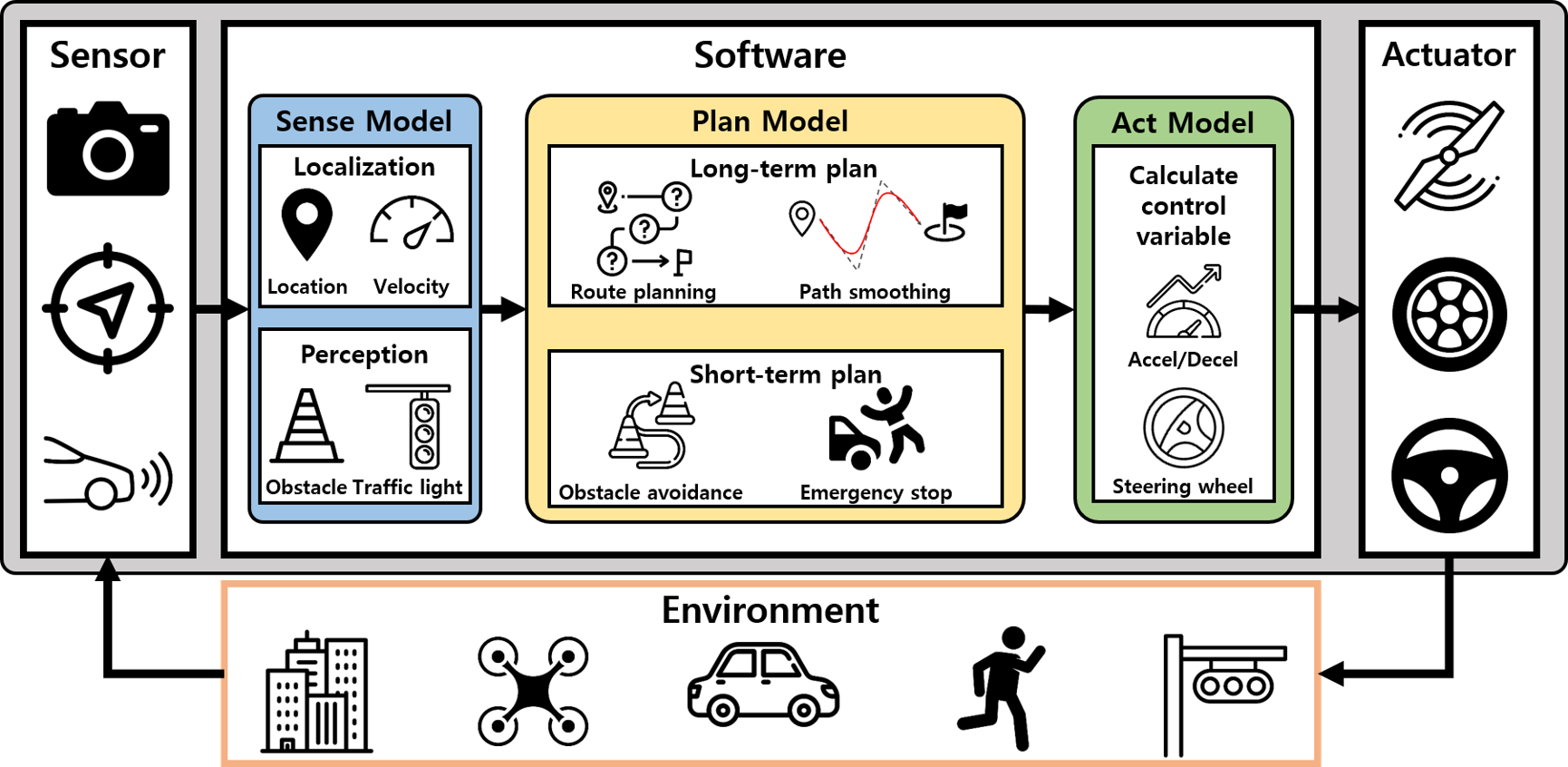}}
    \caption
    {The Sense-Plan-Act model of ADS}
    \label{fig:ads-arch}
    \vspace{-5mm}
\end{figure}

Existing ADS testing frameworks often focus on the Sense model within the Sense-Plan-Act architecture, as illustrated in ~\cref{fig:ads-arch}, by attacking perception through adversarial sensor inputs or simulation~\cite{jing2021too, song2018physical, nassi2020phantom, tu2021exploring, cao2021invisible, hingun2022reap}, or on end-to-end robustness against security attacks~\cite{li2020av, kim2022drivefuzz, zhong2022neural, wan2022too, sun2022lawbreaker}.
Plan model testing has received relatively less attention.
Prior works have generated scenarios from accident databases or predefined templates, but often without consideration of how planner internals interpret those scenarios.
Most treat the ADS as a black box and miss opportunities for white-box guidance.
They often lack root cause attribution and fail to expose logic-specific flaws that propagate over time.

Modern Plan models are implemented as structured control pipelines.
In Baidu Apollo, a widely used open-source Level 4 ADS, planning logic is organized into a scene-stage-task hierarchy.
Each frame begins by classifying the current traffic situation into a scene (e.g., lane following, stopping for a crosswalk), which determines a set of planning stages (e.g., cruise, stop, creep).
These in turn activate task modules such as deciders and optimizers that generate candidate trajectories.
This structure reflects the planner’s internal assumptions about operational context and permissible responses.
It also introduces implicit constraints: only certain transitions between scenes are allowed; certain stages activate only in specific contexts; and individual modules may or may not influence final motion depending on execution flow.

Despite the well-defined structure of these systems, existing testing techniques often fail to leverage it.
Two key challenges arise in testing Plan models.
First, generating test inputs that comprehensively exercise planner behavior is non-trivial.
Prior work typically relies on black-box scenario generation: modifying road geometry, vehicle trajectories, or environmental parameters without knowledge of how such changes affect the planner's internal decision-making.
These approaches often yield redundant or ineffective test cases, failing to trigger deeper branches of planner logic.
Second, identifying the fault localization of the discovered hazardous behavior is challenging.
Planning failures often emerge over multiple frames, making it difficult to isolate the responsible scene, stage, or task.
Without visibility into the planner’s structure or internal states, debugging becomes time-consuming and imprecise.

To address these issues, we propose \sys, a guided fuzzer designed to systematically identify hazardous behaviors within ADS Plan models through end-to-end ADS testing.
We target Baidu Apollo~\cite{apollo} as our evaluation platform.
Baidu Apollo is an open-source level 4 ADS software that provides a complete technology stack, from hardware requirements to core software components necessary for autonomous driving.
Our approach consists of three key components: 1) utilizing diverse driving scenarios generated in accordance with Operational Design Domain (ODD) specifications as inputs to \sys, 2) detecting hazardous behaviors through three bug oracles and novel scoring metrics, and 3) classifying the discovered hazardous behaviors based on their root causes.

Through our implementation of \sys, we identified 520 hazardous behaviors in 20,635 test cases.
To systematically analyze these results, we established bug classification criteria using information from ADS internal messages and simulator data.
Through our classification process, we initially grouped the 520 hazardous behaviors into 156 distinct cases, which were further consolidated into 15 categories through root cause analysis.
The validity of our root cause analysis was confirmed through the successful resolution of several identified issues.
To evaluate the comprehensiveness of our testing approach, we measured both functional coverage and decision coverage, achieving 83.63\% and 63.22\% respectively.
Furthermore, we demonstrated \sys's performance by comparing it with two baseline testing algorithms, showing its ability to generate more critical scenarios within the same time frame.
A demonstrative video of the discovered hazardous behaviors is available at \url{https://sites.google.com/view/safe-planner/home}.

In summary, our contributions are as follows:
\begin{itemize}
    \item We propose a novel approach for driving scenario generation by decomposing scenarios into scene transitions and NPC behaviors, enabling systematic generation of diverse and realistic test cases. We extracted 80 possible scene transitions from Baidu Apollo and generated 32 NPC behavior types based on ISO 34502~\cite{ISO34502:2022}.
    \item We introduce \sys, a testing framework designed to discover hazardous behaviors in Baidu Apollo Plan Model. Through manual analysis of Baidu Apollo's internal structure, \sys generates comprehensive driving scenarios and employs novel scoring metrics to detect hazardous behaviors in the Plan Model.
    \item We demonstrate the effectiveness of \sys through extensive evaluation, discovering 520 hazardous behaviors that were classified into 15 distinct categories based on ADS internal messages and simulator data. We present detailed root cause analysis and case studies, providing insights into failure patterns and their implications.
\end{itemize}

The remainder of the paper is structured as follows.
In \cref{sec:background}, we describe background knowledge relevant to our work.
\cref{sec:approach} outlines the system overview and describes our systematic approach to defining and generating driving scenarios.
In \cref{sec:evaluation}, we present the evaluation results.
\cref{sec:related} reviews related work and \cref{sec:conclusion} concludes.

\section{Background: Plan Model of Baidu Apollo}
\label{sec:background}

The Plan model of Baidu Apollo is critical for generating safe and efficient trajectories for the ego vehicle.
To achieve this, Baidu Apollo employs a hierarchical structure consisting of three layers.
In this hierarchical approach, a scene represents an abstract behavioral context (such as lane following or emergency stop), a stage represents a specific sub-behavior within that scene (such as stopping or creeping), and a task represents the lowest-level computational actions for generating trajectory points.
In Baidu Apollo's architecture, each complete cycle of the control loop is referred to as a frame. 
At every frame, the Plan model of Baidu Apollo selects an appropriate scene corresponding to the perceived data from the Sense model.
Each scene consists of several stages, and each stage executes a series of tasks.
This hierarchical structure is organized to process common trajectory generation tasks at the lowest level, while handling situation-specific behaviors at higher levels.
The following sections provide detailed explanations of the hierarchical structure.

\nibf{Scene} The scene represents the abstracted behavior of the ego vehicle in a specific situation.
In other words, the ego vehicle selects an appropriate scene to deal with the current environment, which is the combination of the state of the ego vehicle and static entities around the ego vehicle including the traffic light and road signs.
Specifically, the state of the ego vehicle includes the current location, destination, and signals from the Human Machine Interface (HMI) while static entities include the traffic light, road signs, and road information.
To illustrate this concept, the ego vehicle operated by Baidu Apollo will select the traffic light unprotected right turn scene when it needs to make a right turn at a nearby intersection to reach its destination.
Alternatively, if the driver sends an emergency stop signal through the HMI, the ego vehicle will select the emergency stop scene regardless of its current location and operate accordingly.
Following this approach, Baidu Apollo defines 14 distinct scenes and continuously selects the appropriate scene at each frame during operation.

\nibf{Stage} At the intermediate level of the hierarchical structure, the stage corresponds to sub-behaviors within the scene.
It subdivides the behavior of each scene and is executed in sequence to complete the scene.
For example, in the traffic light unprotected right turn scene, the stop stage is first executed to stop at the stop line before entering the intersection.
Then, the creep stage is executed to enter into the intersection at an appropriate speed.
Finally, the cruise stage is executed to pass through the intersection.
As shown in this example, each scene consists of multiple stages that are executed in a predetermined sequence to achieve the desired behavior.

\nibf{Task} Unlike the higher levels that focuses on the abstracted behavior of the ego vehicle, the task focuses on the common and low-level behavior, such as generating concrete trajectory points.
Because of this design, most stages consist of a sequence of commonly shared tasks.
These common task sequences generate safe and efficient trajectory points based on the environment data from the Sense model and scene- or stage-specific information transferred from the higher level.
Also, tasks are divided into two types, deciders and optimizers.
Deciders determine low-level behavior to generate safe trajectory points, and optimizers optimize the trajectory points to improve efficiency.
Deciders and optimizers work cooperatively to generate trajectory points that are both safe and efficient.

\section{Approach}
\label{sec:approach}

\subsection{System Overview}
\label{subsec:overview}

\begin{figure*}[t]
    \centerline{\includegraphics[width=0.9\linewidth]{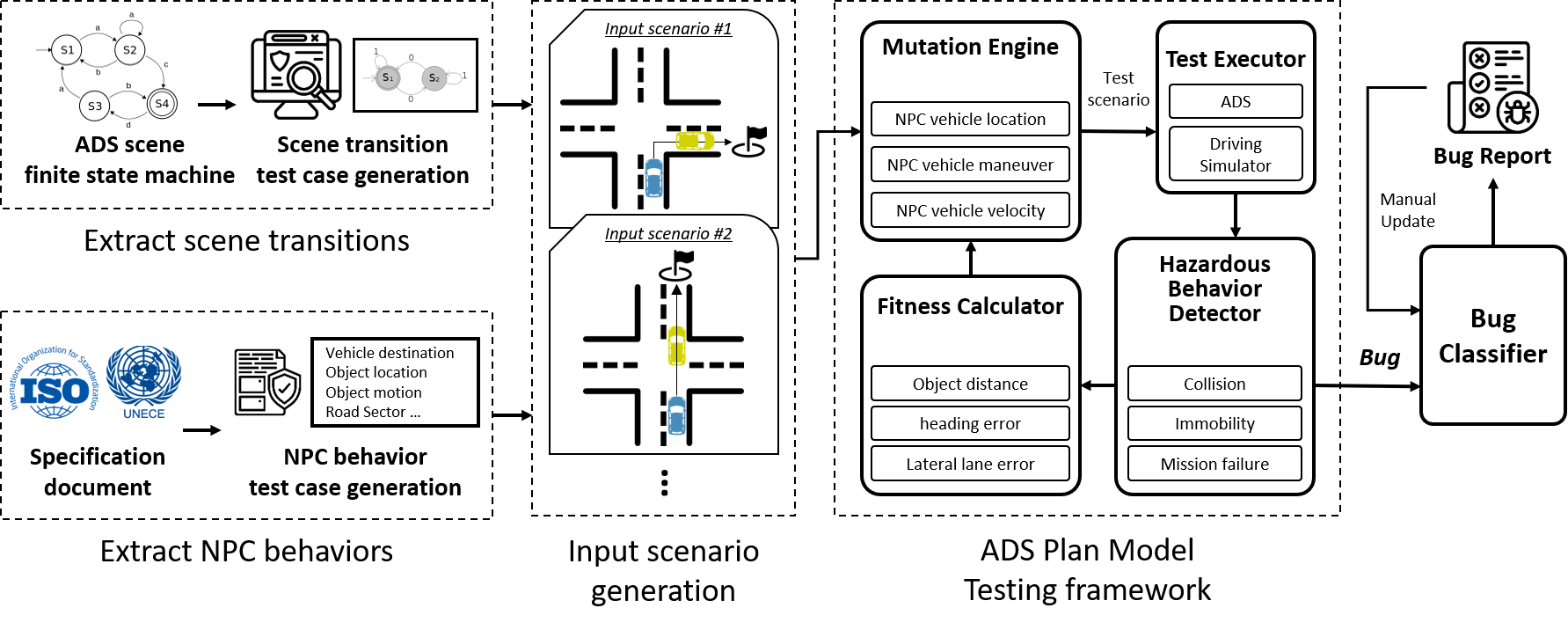}}
    \vspace{-2mm}
    \caption{Overview of \sys}
    \label{fig:overview}
    \vspace{-3mm}
\end{figure*}

We propose \sys, a testing framework designed to systematically discover functional insufficiencies within the Plan model of an ADS. 
\sys aims to detect hazardous behavior in the Plan model by taking various driving scenarios as input.
It evaluates each scenario and detect the hazardous behaviors and clusters them.
\sys, as depicted in~\cref{fig:overview}, consists of two major components: Input scenario generation and Plan model testing framework.
The input scenarios generation is the initial step in the \sys framework.

Driving scenarios serve as input for the testing framework and consist of two components: a series of \textit{scene transitions} and an \textit{NPC behavior}.
Scene transition means the change from one scene to another, and each scene represents a specific circumstance faced by the ego vehicle.
On the other hand, the NPC behavior defines the location and maneuver of non-ego vehicles within the scenario.
By generating different scene transitions and corresponding NPC behavior, we can create diverse input scenarios (see \cref{subsec:scenario_generation} for more details).

Secondly, we utilize these input scenarios to identify hazardous behaviors of the ADS Plan model.
\sys 1) uses the scene as a seed for fuzzer, 2) mutates NPC behavior and combines with a scene to generate a test scenario, 3) detects bugs using three bug oracles: collision, immobility, and mission failure, and 4) evaluate the test scenario with new score metric associated with the bug oracles in order to guide toward buggy directions (see \cref{subsec:mutation,subsec:fitness,subsec:detector}).

\begin{algorithm}[t]
\caption{\sys algorithm}
\label{alg:system_alg}
\DontPrintSemicolon
\KwIn{A set of seed scenarios $\mathbf{S}$}
\KwOut{A set of hazardous scenario $\mathbf{H}$}
\Parameter{Population size $P$, Number of generations $N$}
$\mathbf{H} \gets \{\}$ \;
\For{$s_{seed} \in \mathbf{S}$} {
  $\mathbf{Q} \gets \{s_{seed},\cdots,s_{seed}\}$ \;
  \For {$i \in \{1,\cdots,N\}$} {
    $\mathbf{F} \gets \{\}$ \;
    $\mathbf{Q^\prime} \gets \delta(\mathbf{Q})$ \tcp{Mutation}
    \For{$\hat{s} \in \mathbf{Q^\prime}$} {
      $f(\hat{s}), O(\hat{s}) \gets \texttt{SimExecution}(\hat{s})$ \tcp{Fitness Calculator}
      $\mathbf{F} \gets \mathbf{F} \cup \{f(\hat{s})\}$ \;
      \If{$O(\hat{s}) =$ FAILURE \tcp{Hazardous behavior check}} {
        $\mathbf{H} \gets \mathbf{H} \cup \{\hat{s}\}$
      }
    }
    $\mathbf{Q} \gets \texttt{Selection}(\mathbf{Q^\prime}, \mathbf{F})$ \tcp{Feedback}
  }
}
\Return $\mathbf{H}$ \;
\end{algorithm}

\cref{alg:system_alg} presents the main algorithmic procedure of \sys.
\sys receives a set of initial seed scenario $S$ as inputs and outputs a set of a hazardous scenario $H$.
Parameters $P$ and $N$ can be adjusted to configure the population size and number of generations, respectively.
The algorithm begins by creating an initial population with the seed scenario $s_{seed}$ selected from the seed scenario set $S$ (Lines 2-3).
In each iteration, the algorithm first generates the offspring by mutating each scenario in the
population Q (Line 6).
Each new scenario $\hat{s}$ is first executed in the simulation platform consisting of a simulator and the ADS under test and evaluates the fitness score of each scenario (Line 8).
If the output of the scenario execution failed to pass the hazardous behavior check, the identified hazardous scenario is added to the hazardous scenario set $H$ (Lines 10-12).
If no hazardous scenario is detected, the scenario selection picks the top scenario from the
population $Q$ based on their fitness scores and repeats from line 4 (Line 13).
The algorithm ends by returning detected hazardous scenario set $H$ (Line 14).

In the following sections, we introduce the key components of \sys: the Input Scenario Generation, the Mutation Engine ($\delta$), the Fitness Calculator ($f$), and the Hazardous Behavior Check ($O$).

\subsection{Driving Scenario Generation}
\label{subsec:scenario_generation}

In this section, we present a systematic approach to define and categorize driving scenarios.
We outline the process of classifying two key components: 1) scene transitions and 2) NPC behaviors.

\nibf{Scenario Definition}
Here, we first define a driving scenario.
A driving scenario is defined as a series of specific situations encountered by the ego vehicle while navigating from one location to another.
Given that ADS processes these situations in discrete time intervals, a scenario can be formally expressed as a temporal sequence of scenes.

\vspace{-3mm}
\begin{equation}
\vspace{-1mm}
    scenario = \{scene_1, scene_2, \cdots, scene_n\}
\end{equation}
Here, a scene represents a snapshot that captures the states of all entities, including the ego vehicle, dynamic entities, and static entities.
Dynamic entities encompass objects whose states change during scenario progression, such as NPC vehicles, pedestrians, and motorcycles.
In contrast, static entities remain unchanged throughout the scenario, including roads, traffic signs, and other fixed infrastructure elements.
This can be formally expressed as:

\vspace{-3mm}
\begin{equation}
\vspace{-1mm}
scene = \{V_{ego}, E_{static}, E_{dynamic}\}
\end{equation}
where $V_{ego}$ represents the ego vehicle state, $E_{static}$ represents the set of static entities, and $E_{dynamic}$ represents the set of dynamic entities.
ADS abstract and classify scenes to 1) reduce the complexity of scene representation and 2) interpret and respond effectively for decision making.
To manage complexity in scene abstraction and classification, scenes are abstracted to include only the ego vehicle and static entities, excluding dynamic entities which would exponentially increase the number of possible cases.
The abstracted scenes, denoted as $s_{abstract} = {V_{ego}, E_{static}}$, captures fundamental driving contexts such as turning right at an intersection or stopping at a stop sign, while maintaining a tractable set of scenarios.

Based on this decomposition, a scenario can be decomposed into two components: a series of abstracted scenes and a series of dynamic entity states ($E_{dynamic}$).
The change in $s_{abstract}$ can be defined as scene transitions, $T_i = \langle s_{i}, s_{i+1} \rangle$, and the change in dynamic entity states can be defined as dynamic entity behavior, i.e., NPC behavior $B_i = \langle E_{dynamic,i}, E_{dynamic,i+1} \rangle$.
In summary, we can abstractly define a scenario as the combination of scene transitions and NPC behavior:

\vspace{-3mm}
\begin{equation}
\vspace{-1mm}
scenario = \{(T_i, B_i)~|~i = 1, 2, \ldots, n-1\}
\end{equation}

\begin{table}[t]
\scriptsize
\centering

\setlength\tabcolsep{0.030cm}
\def\arraystretch{0.75}

\caption{Scene transitions extracted from Baidu Apollo}
\label{tab:scene_transition}

\begin{threeparttable}
\begin{tabular}{@{}l|cccccccccccc@{}}
\toprule

& LF & PO & TLP & TLULT & TLURT & SSU & YS & BIU & VP & PG & EPO & ES \\
\midrule

LF & \CIRCLE & \Circle & \CIRCLE & \CIRCLE & \CIRCLE & \CIRCLE & \Circle & \CIRCLE & \Circle & \Circle & \CIRCLE & \CIRCLE \\
\midrule
PO & \Circle & \Circle & \Circle & \Circle & \Circle & \Circle & \Circle & \Circle & \Circle & \Circle & \Circle & \Circle \\
\midrule
TLP & \CIRCLE & -- & \CIRCLE & -- & -- & -- & -- & -- & \Circle & \Circle & \CIRCLE & \CIRCLE \\
\midrule
TLULT & \CIRCLE & -- & -- & \CIRCLE & -- & -- & -- & -- & \Circle & \Circle & \CIRCLE & \CIRCLE \\
\midrule
TLURT & \CIRCLE & -- & -- & -- & \CIRCLE & -- & -- & -- & \Circle & \Circle & \CIRCLE & \CIRCLE \\
\midrule
SSU & \CIRCLE & -- & -- & -- & -- & \CIRCLE & -- & -- & \Circle & \Circle & \CIRCLE & \CIRCLE \\
\midrule
YS & \Circle & -- & -- & -- & -- & -- & \Circle & -- & \Circle & \Circle & \Circle & \Circle \\
\midrule
BIU & \CIRCLE & -- & -- & -- & -- & -- & -- & \CIRCLE & \Circle & \Circle & \CIRCLE & \CIRCLE \\
\midrule
VP & \Circle & -- & -- & -- & -- & -- & -- & -- & \Circle & \Circle & \Circle & \Circle \\
\midrule
PG & \Circle & -- & \Circle & \Circle & \Circle & \Circle & \Circle & \Circle & \Circle & \Circle & \Circle & \Circle \\
\midrule
EPO & -- & -- & -- & -- & -- & -- & -- & -- & -- & \Circle & \CIRCLE & -- \\
\midrule
ES & -- & -- & -- & -- & -- & -- & -- & -- & -- & \Circle & -- & \CIRCLE \\
\bottomrule
\end{tabular}
\begin{tablenotes}
  \item [$\bigstar$] Each cell depicts the possible scene transitions from rows to columns. For instance, the third cell in PO row indicates that a scene transition from PO to TLP is possible.
  \item [$\dag$] \CIRCLE~and \Circle~in each cell indicate whether the scene transition was used or not used in the evaluation, respectively.
\end{tablenotes}
\end{threeparttable}
\vspace{-4mm}
\end{table}

\begin{table*}[t]
\scriptsize
\centering

\setlength\tabcolsep{0.10cm}
\def\arraystretch{1}

\caption{Scene list defined in Baidu Apollo}
\label{tab:scene_list}

\begin{threeparttable}
\begin{tabular}{@{}ccll@{}}
\toprule

Scene \# & Scene abbr. & Scene name & Description\\
\midrule

S1 & LF & LANE\_FOLLOW & Driving the straightforward lane \\
S2 & PO & PULL\_OVER & Stop at the side of the road when arrived at destination \\
S3 & TLP & TRAFFIC\_LIGHT\_PROTECTED & Go straight at intersection where there are traffic lights \\
S4 & TLULT & TRAFFIC\_LIGHT\_UNPROTECTED\_LEFT\_TURN & Turn left at intersection where there are traffic lights \\
S5 & TLURT & TRAFFIC\_LIGHT\_UNPROTECTED\_RIGHT\_TURN & Turn right at intersection where there are traffic lights \\
S6 & SSU & STOP\_SIGN\_UNPROTECTED & Stop near a stop sign and resume the mission \\
S7 & YS & YIELD\_SIGN & Send the other vehicle first near the yield sign \\
S8 & BIU & BARE\_INTERSECTION\_UNPROTECTED & Driving the intersection without traffic lights and traffic signs \\
S9 & VP & VALET\_PARKING & Park at the pre-defined parking spot \\
S10 & PG & PARK\_AND\_GO & Returning to the road from a non-city road, such as the side of a road \\
S11 & EPO & EMERGENCY\_PULL\_OVER & Move to the side of the road and stop \\
S12 & ES & EMERGENCY\_STOP & Stop in the current lane \\
S13 & DT & DEADEND\_TURNAROUND & At a dead end, turn around and go back \\
S14 & SSP & STOP\_SIGN\_PROTECTED\dag & Stop at the side of the road when arrived at destination \\

\bottomrule
\end{tabular}

\begin{tablenotes}
 \item [\dag] Determined as SSU or SSP according to flag value $stop\_sign\_all\_way$
\end{tablenotes}

\end{threeparttable}
\vspace{-3mm}
\end{table*}

\nibf{Scene Transition}
There is no universally defined standard for dividing scenes in the context of ADS.
As a result, each ADS may employ its own approach to segment and define scenes, which can lead to variations in how the scenes are divided.
Hence, comprehensively defining all possible scenes and constructing a corresponding FSM for ADS is challenging.

To address this challenge, we employ a method that involves collecting abstracted scenes and scene transitions from the target system, Baidu Apollo v7.0.0.
Baidu Apollo consists of 14 abstracted scenes, which include primary lane following, traffic lights, bare crossings, traffic signs, emergencies, parking, and unexpected situations~(\cref{tab:scene_list}).
These abstracted scenes offer a total of 80 possible scene transitions, including the option to maintain the current state.

In order to create a comprehensive test scenario, we need to be able to test all possible scene transitions.
However, the implementation of scene transitions for testing presents certain constraints due to differences between ADS maps and simulator maps.
While ADS utilize High-Definition (HD) maps that incorporate detailed information essential for autonomous driving, these maps differ significantly from simulator maps in their structure and content.
HD maps specifically store condensed information including lane widths, lane connectivity, and traffic signal locations, whereas simulator maps contain comprehensive object definitions for rendering and physics simulation.
Typically, HD maps are generated through a conversion process where developers annotate simulator maps with necessary autonomous driving information.
However, this process can lead to testing limitations for certain scene transitions.
These limitations manifest in two primary ways: 1) missing annotations in the simulator map, or 2) improper exportation of annotated information to the HD map.
While the latter can be addressed through modifications, the former requires complex annotation processes, leading us to exclude scene transitions related to YS, VP, PG, and DT from our testing scope.
Furthermore, we excluded PO and SSP scene transitions that would require configuration modifications for testing.
Consequently, as shown in~\cref{tab:scene_transition}, we identified 30 scene transitions suitable for testing from the original 80 possibilities, which are indicated by filled circles.

\begin{figure}[t]
    \centering
    \begin{subfigure}{0.46\columnwidth}
        \centering
        \includegraphics[height=3.5cm]{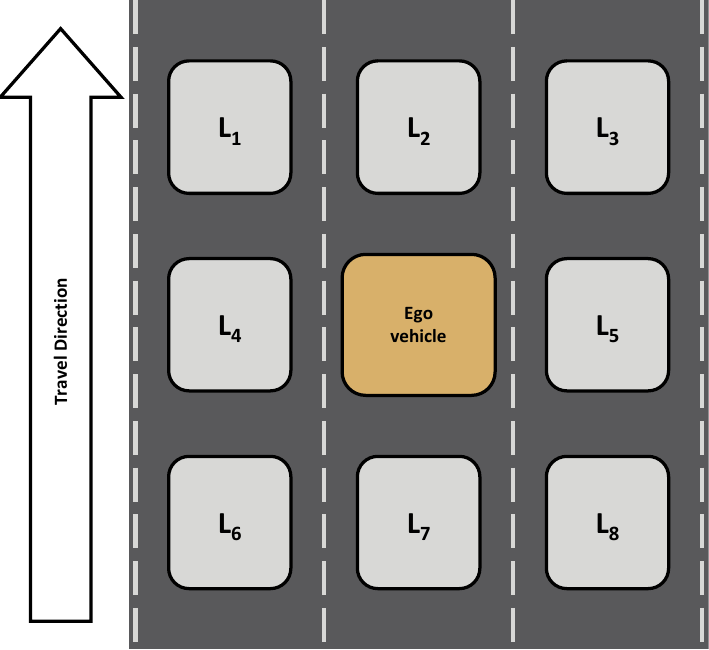}
        \caption{NPC vehicle locations}
        \label{subfig:npc_location}
    \end{subfigure}
    \begin{subfigure}{0.46\columnwidth}
        \centering
        \includegraphics[height=3.5cm]{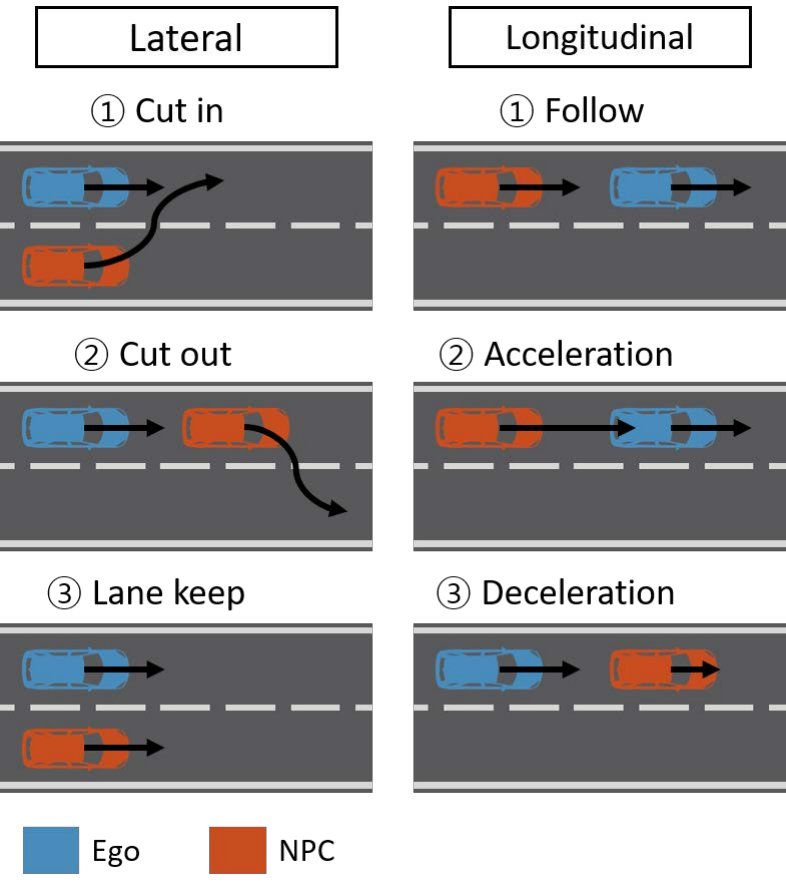}
        \caption{NPC vehicle maneuvers}
        \label{subfig:npc_maneuver}
    \end{subfigure}
    \caption{Location and maneuver of NPC behavior}
    \vspace{-2mm}
    \label{fig:location_map}
\end{figure}

\nibf{NPC Behavior}
The behavior of the NPC vehicle is particularly important as it directly influences the behavior of the ego vehicle.
Therefore, it is crucial to define the behavior of the NPC vehicle systematically to generate the non-hazardous scenario.
In order to accomplish this, we define the NPC vehicle as the same directional vehicle movement because the most commonly the ego vehicle encounter is the NPC vehicle traveling in the same direction on the road.
For simplicity, we only consider one NPC vehicle.

The behavior of a NPC can be described as a combination of location and maneuver as mentioned in the ISO 34502~\cite{ISO34502:2022}. 
The maneuver of a NPC can be described as a combination of longitudinal maneuver $M_{Lon}$ and lateral maneuver $M_{Lat}$.
For simplicity, we treat those two types of maneuvers as the control inputs of the bicycle model~\cite{kong2015kinematic}: acceleration $a(t)$ and steering angle $\psi(t)$. 

\vspace{-5mm}
\begin{equation*}
\vspace{-1mm}
\begin{split}
    Maneuver = (M_{Lon}, M_{Lat}) = (a(t), \psi(t))
\end{split}
\end{equation*}

For the longitudinal maneuver $M_{Lon}$, the acceleration can take on three values: zero, positive, and negative, which respectively correspond to following, acceleration, and deceleration behaviors.
On the other hand, the lateral maneuver $M_{Lat}$ can be classified into cut-in, cut-out, and lane-keeping behaviors based on the direction of $\psi(t)$ relative to the ego vehicle, as shown in~\cref{subfig:npc_maneuver}.
By combining $M_{Lon}$ and $M_{Lat}$, nine fundamental maneuvers are generated for the NPC vehicle.

Then, this maneuver is paired with the location of the NPC with respect to the ego vehicle. As depicted in~\cref{subfig:npc_location}, there are 8 possible NPC locations ($L_1 \sim L_8$) around the ego vehicle. This categorization enables us to determine the relative position of the NPC with respect to the ego vehicle.

We focus on the NPC behavior that may interfere with the trajectory of the ego vehicle.
For example, we consider that deceleration of vehicles at certain locations may not pose a threat if the distance between vehicles increases ($L_6$, $L_7$, $L_8$).
Therefore, we exclude motions that result in an increase in the relative distance from the input scenario.
The circles in the~\cref{tab:input_surrounding} represent combinations of surrounding vehicle locations and motions that may pose a threat to the ego vehicle.
Based on these combinations, we consider a total of 32 types of possible behaviors in the input scenario.
The NPC behavior also includes velocity and distance (between the ego and NPC vehicle) values.

\begin{table}[t]
\scriptsize
\centering

\setlength\tabcolsep{0.030cm}
\def\arraystretch{0.75}

\caption{SOTIF related NPC behavior types}
\label{tab:input_surrounding}

\begin{threeparttable}
\begin{tabular}{@{}lcccccccccc@{}}
\toprule

\multirow{2}{*}{
\begin{tabular}{c} Surrounding \\ vehicle location \end{tabular}
}
& \multicolumn{3}{c}{\begin{tabular}[c]{@{}c@{}} Cut in\end{tabular}} & \multicolumn{3}{c}{\begin{tabular}[c]{@{}c@{}} Cut out\end{tabular}} & \multicolumn{3}{c}{\begin{tabular}[c]{@{}c@{}} Lane keep\end{tabular}} \\
\cmidrule(lr){2-4}
\cmidrule(lr){5-7}
\cmidrule(lr){8-10}

& \multicolumn{1}{c}{Follow} & \multicolumn{1}{c}{Accel} & \multicolumn{1}{c}{Decel} & \multicolumn{1}{c}{Follow} & \multicolumn{1}{c}{Accel} & \multicolumn{1}{c}{Decel} & \multicolumn{1}{c}{Follow} & \multicolumn{1}{c}{Accel} & \multicolumn{1}{c}{Decel} \\
\midrule

\multicolumn{1}{l|}{\textbf{L$_1$} (North-West)} & \Circle & -- & \Circle & -- & -- & -- & \Circle & -- & \Circle \\
\midrule
\multicolumn{1}{l|}{\textbf{L$_2$} (North)} & -- & -- & -- & \Circle & -- & \Circle & \Circle & -- & \Circle \\
\midrule
\multicolumn{1}{l|}{\textbf{L$_3$} (North-East)} & \Circle & -- & \Circle & -- & -- & -- & \Circle & -- & \Circle \\
\midrule
\multicolumn{1}{l|}{\textbf{L$_4$} (West)} & \Circle & \Circle & \Circle & -- & -- & -- & \Circle & -- & -- \\
\midrule
\multicolumn{1}{l|}{\textbf{L$_5$} (East)} & \Circle & \Circle & \Circle & -- & -- & -- & \Circle & -- & -- \\
\midrule
\multicolumn{1}{l|}{\textbf{L$_6$} (South-West)} & \Circle & \Circle & -- & -- & -- & -- & \Circle & \Circle & -- \\
\midrule
\multicolumn{1}{l|}{\textbf{L$_7$} (South)} & -- & -- & -- & \Circle & \Circle & -- & \Circle & \Circle & -- \\
\midrule
\multicolumn{1}{l|}{\textbf{L$_8$} (South-East)} & \Circle & \Circle & -- & -- & -- & -- & \Circle & \Circle & -- \\

\bottomrule
\end{tabular}


\end{threeparttable}
\vspace{-3mm}
\end{table}

\nibf{Preprocessing Step}
We instrumented various parameters for our testing framework, including Baidu Apollo's internal planning variables, simulator data for both ego and NPC vehicles, coverage metrics, feedback scores, and other information required for bug classification.
Note that we do not modify any parameters of the target system.

To establish our initial seed set, we generated seed scenarios by combining each possible scene transition with corresponding NPC behaviors, creating one seed for each unique combination.
For each scene transition, we randomly selected the ego vehicle's initial state and destination state to ensure diverse starting conditions.
Through iterative testing and analysis, we subsequently augmented the seed set by incorporating additional seeds that increase code coverage, thereby enhancing the comprehensiveness of our testing framework.

\subsection{Mutation Engine}
\label{subsec:mutation}
The mutation engine in \sys is responsible for mutating the NPC behavior in input scenarios, which includes the NPC's location, maneuver, and initial speed.

\nibf{NPC location mutation.}
When generating NPC vehicles, \sys specifically focuses on vehicles that are traveling in the same direction as the ego vehicle.
This approach is based on the understanding that the ego vehicle generally avoids interactions with vehicles in the opposite lane, except in cases of single directional roads, bare intersections, and intentional centerline violations.
To ensure the generation of NPC vehicles in the appropriate regions, the mutation engine examines the presence of neighboring forward lanes on both the left and right sides of the ego vehicle's location. 
Once the examination is complete, the mutation engine randomly selects one region from L$_1$ to L$_8$.

For example, if only the left side of the ego vehicle has a forward lane, the NPC vehicles will be generated exclusively in the L$_1$, L$_2$, L$_4$, L$_6$, and L$_7$ regions.
After determining the region, the mutation engine randomly chooses the location of the NPC vehicle within the selected region.
It is important to note that the location is carefully selected within a configurable range around the initial location of the ego vehicle.
This ensures that the NPC vehicle is always generated within the bounds of interaction with the ego vehicle.

\nibf{NPC velocity mutation.}
In our evaluation, we randomly select the initial speed of the NPC vehicle within the range of 0 to the road speed limit, which is 11.18~m/s (40~km/h) on the map we used. The initial speed of the NPC vehicle has a notable impact on the behavior of the ego vehicle. When an NPC vehicle cuts into the ego vehicle's lane from an adjacent lane, the ego vehicle will yield and slow down if the NPC vehicle's speed is high. Conversely, if the NPC vehicle's speed is low, the ego vehicle may choose to overtake. Therefore, conducting tests with varying initial speeds is crucial to fully comprehend the impact of this parameter.

\subsection{Fitness Calculator}
\label{subsec:fitness}
Our fuzzer, powered by the GA, employs guided fuzzing to identify bugs using a fitness score that guides the search toward specific hazardous behaviors.
We have developed a score metric that targets Dynamic Driving Tasks (DDTs) which may lead to hazardous behavior during scenarios~\cite{ISO21448:2022}.
The score metric focuses on DDTs, which encompass operational and tactical functions for real-time vehicle operation, including object and event detection and response (OEDR) and longitudinal and lateral vehicle motion control~\cite{ISO22736:2021}.

Our fitness calculator evaluates driving quality by considering three key factors: distance to obstacle vehicles, lateral distance from lane center, and heading angle difference relative to the lane.

\nibf{Distance to NPC vehicle.}
While the relationship between the distance and hazardous behavior may not be linear, a closer proximity between the vehicles increases the likelihood of an accident.
Therefore, the average distance between the ego vehicle and the NPC vehicle during the scenario, denoted as $\vartriangle d_{obs,avg}$, serves as an argument of our scoring metric.

\vspace{-1mm}
\begin{equation}
\vspace{-1mm}
    \vartriangle d_{obs,avg} = \frac{\sum_{t=1}^T \vartriangle d_{obs}[t]}{T}
\end{equation}
where $t$ represents each time frame, $\vartriangle d_{obs}[t]$ represents the distance at time frame $t$, and $T$ is the total number of time frames in the scenario.

\nibf{Lateral distance from lane.}
This measures the distance between the ego vehicle and target lane center to assess collision risk during lateral maneuvers.
Precise lateral movement is essential to avoid collisions with vehicles in neighboring lanes, particularly during lane changes.
The average lateral distance $\vartriangle d_{lat,avg}$ is calculated as:

\vspace{-1mm}
\begin{equation}
\vspace{-1mm}
    \vartriangle d_{lat,avg} = \frac{\sum_{t=1}^T \vartriangle d_{lat}[t]}{T}
\end{equation}

\nibf{Heading angle difference from lane.}
When the angle between the heading of the ego vehicle and the target lane significantly increases, the vehicle is more likely to deviate further from the center of the lane.
Consequently, as the heading angle difference increases, the ego vehicle faces challenges in maintaining its position within the lane, which can result in drifting into neighboring lanes or experiencing side-to-side movements.
We calculate the average heading angle difference $\vartriangle\theta_{avg}$:


\vspace{-1mm}
\begin{equation}
\vspace{-1mm}
    \vartriangle\theta_{avg} = \frac{\sum_{t=1}^T \vartriangle\theta[t]}{T}
\end{equation}

\nibf{Overall metric.}
To efficiently detect functional insufficiencies that may occur in DDTs across various scenarios, our fitness calculator employs a weighted sum metric.
Each of the three metrics described above is assigned an individual weight.
These weights were derived through experimental tuning, aiming to strike a balance between achieving sufficient coverage and maintaining computational efficiency.

\vspace{-4mm}
\begin{equation}
    score = \omega_0\cdot\vartriangle d_{obs,avg} + \omega_1\cdot\vartriangle d_{lat,avg} + \omega_2\cdot\vartriangle\theta_{avg}
\end{equation}

\subsection{Hazardous Behavior Detector}
\label{subsec:detector}
In \sys, failure of the ADS refers to the inability to reach the destination without hazardous situations and to the failure of emergency missions. 
Our hazardous behavior detector aims to identify failures of Plan model by monitoring events directly related to human safety or time constraints, collision, immobility, and mission failure.

\nibf{Collision.}
One of the most dangerous scenarios that can lead to severe harm to human passengers is a collision.
The detector identifies when the ego vehicle physically contacts NPC vehicles or objects in the simulator.
This differs from collision decisions made by the ADS Plan model—cases where the ADS determines a collision without actual impact and stops are categorized as immobility.

\nibf{Immobility.}
Immobility is an undesired event that occurs when the ego vehicle remains motionless for an extended period of time, unless it is intentional parking.
Although the ego vehicle may come to a temporary stop during driving when necessary, such as recognizing a stop sign or encountering an obstacle in front, it is expected that this stationary condition is temporary. 
This occurs when the ego vehicle remains motionless for an extended period beyond intentional stops (e.g., stop signs, temporary obstacles).
We set a time limit $t_{stop}$; if the ego vehicle does not move continuously during this period, immobility is detected.

\nibf{Mission failure.}
Mission failure, within the context of the Plan model in an ADS, refers to the inability to successfully complete the given mission without encountering hazardous situations.
This encompasses two scenarios: 1) failure to reach the destination within time threshold $t_{dest}$, and 2) failure to complete emergency missions.
For emergency stop missions, failure is detected if the ego vehicle does not stop ($v_{ego} > v_{thres}$) at $T_{emer}$.
For emergency pullover missions, failure occurs if the vehicle does not stop in the rightmost lane at $T_{emer}$.
\section{Evaluation}
\label{sec:evaluation}

In this section, we empirically evaluate \sys's capability to detect hazardous behaviors in ADS Plan model.
We address the following research questions:

\nibf{RQ1:} Can \sys effectively discover diverse hazardous behaviors in the ADS Plan model? \\
\nibf{RQ2:} Does \sys provide comprehensive coverage of diverse scenarios for testing ADS Plan model? \\
\nibf{RQ3:} Can \sys efficiently discover diverse hazardous behaviors in the ADS Plan model?

To answer these research questions, we conduct experiments using the following settings:

\nibf{Target ADS.}
We evaluated Baidu Apollo version 7.0~\cite{apollo} as a target ADS system and SORA-SVL simulator.

\nibf{Test scenario generation.}
We generated 712 seed scenarios by combining 30 scene transitions with possible NPC behavior types.
These seed scenarios were then expanded into 20,635 test cases.
To ensure non-hazardous initial conditions, we configured NPC vehicles' starting positions and velocities to maintain appropriate safety distances and speed levels.

\nibf{Testing Setup and Evaluation Metric.}
In evaluations, we used a PC running Ubuntu 20.04, equipped with an Intel i9-12900K CPU, an NVIDIA GeForce RTX 3090 Ti graphics card, and 96GB RAM.
We use the number of \textit{unique hazardous behaviors} as an evaluation metric, where a distinct bug is defined as having different fault localization.

\begin{table*}[t]
\scriptsize
\centering

\setlength\tabcolsep{0.030cm}
\def\arraystretch{0.75}

\caption{Summary of detected hazardous behaviors in Baidu Apollo}
\label{tab:bug_case}

\begin{threeparttable}
\begin{tabular}{@{}cllc@{}}
\toprule

\multirow{2}{*}{\begin{tabular}{c}Bug ID\end{tabular}} & \multirow{2}{*}{\begin{tabular}{c}Bug Description\end{tabular}} & \multirow{2}{*}{\begin{tabular}{c}Triggering Condition\end{tabular}} & \multirow{2}{*}{\begin{tabular}{c}\#HB\end{tabular}} \\
& & & \\

\midrule

01 & Lane borrow immobile/collision & bug oracle = C/I $\cap$ lane borrow decision = True & 68 \\
02 & Nudge immobile/collision &  bug oracle = C/I $\cap$ object nudge decision = True & 23 \\
03 & NPC vehicle Cut-in collision & bug oracle = C $\cap$ lane borrow = False $\cap$ nudge = False $\cap$ npc motion = Cut-in & 20 \\
04 & Stopped NPC vehicle collision & bug oracle = C $\cap$ lane borrow = False $\cap$ nudge = False $\cap$ npc velocity = 0 & 2 \\
05 & Rerouting \& stop sign & bug oracle = I $\cap$ Rerouting = True $\cap$ scene decision = False $\cap$ error msg = (E1 or E2)  & 48 \\
06 & Rerouting \& intersection turn failure & bug oracle = I $\cap$ Rerouting = True $\cap$ error msg = (E1 or E2)  & 50 \\
07 & Intersection turn failure & bug oracle = I/M $\cap$ Rerouting = False $\cap$ error msg = (E1 or E2) & 88 \\
08 & Immobile due to front vehicle & bug oracle = I $\cap$ Rerouting = False $\cap$ (error msg = E3 or stop\_reason\_code = S1) & 30 \\
09 & Immobile during stop sign scene & bug oracle = I $\cap$ Rerouting = False $\cap$ stop\_reason\_code = S2 $\cap$ scene\_end = stop sign & 25 \\
10 & Immobile traffic intersection & bug oracle = I $\cap$ Rerouting = False $\cap$ stop\_reason\_code = S2 $\cap$ scene\_end = Left Turn & 3 \\
11 & Emergency stop failure & bug oracle = M $\cap$ scene\_end = ES $\cap$  ego velocity != 0 & 54 \\
12 & Pull over target change constantly & bug oracle = M $\cap$ scene\_end = EPO $\cap$ pull over position change = True & 65 \\
13 & Pull over failure due to obstacle & bug oracle = M $\cap$ scene\_end = EPO $\cap$ pull over position change = False $\cap$ (error msg = E3 or stop\_reason\_code = S1) & 13 \\
14 & Pull over failure at intersection & bug oracle = M $\cap$ scene\_end = EPO $\cap$ (scene\_start = traffic or bare intersection or stop sign) & 25 \\
15 & Destination reach failure & bug oracle = M $\cap$ scene\_end != (ES and EPO) & 6 \\

\bottomrule
\end{tabular}

\begin{tablenotes}
  \item [] \textbf{I}: Immobility / \textbf{C}: Collision / \textbf{F}: Mission failure
  \item [] \textbf{E1}: planner failed to make a driving plan / \textbf{E2}: Failed to init reference line info.
 / \textbf{E3}: Found collision with obstacle
 \item [] \textbf{S1}: STOP\_REASON\_OBSTACLE / \textbf{S2}: STOP\_REASON\_REFERENCE\_END
 \item [] \#HB stands for the number of hazardous behaviors
\end{tablenotes}
\vspace{-4mm}
\end{threeparttable}
\end{table*}

\subsection{RQ1: Results of Discovering Hazardous Behaviors}
\subsubsection{Total Hazardous Behaviors Results}
\cref{tab:bug_case} summarizes all identified hazardous behaviors.
\sys detected 520 hazardous behaviors among 20,635 test cases.
To verify the diversity of discovered hazardous behaviors, we manually established classification criteria based on data obtained from both ADS internal messages and simulator during test case execution.
These criteria were developed through iterative manual analysis to effectively support fault localization.
Our classification considers bug oracle, ego/NPC vehicle states, rerouting occurrence, scene decision failure, planning status error messages, planning internal messages (such as lane borrow decision, object nudge decision, and stop reason), and test case start/end scenes.

Based on our classification criteria, we initially grouped the 520 hazardous behaviors into 156 distinct cases.
Through detailed analysis, we further consolidated these cases by identifying common fault localization, ultimately resulting in 15 unique hazardous behaviors.
The description, triggering conditions, and the number of hazardous behaviors discovered during evaluation for each unique hazardous behavior are also listed in \cref{tab:bug_case}.

\subsubsection{Case Study}

\begin{figure}[t]
    \centering
    \begin{subfigure}[b]{0.24\textwidth}
        \centering
        \includegraphics[height=3.3cm]{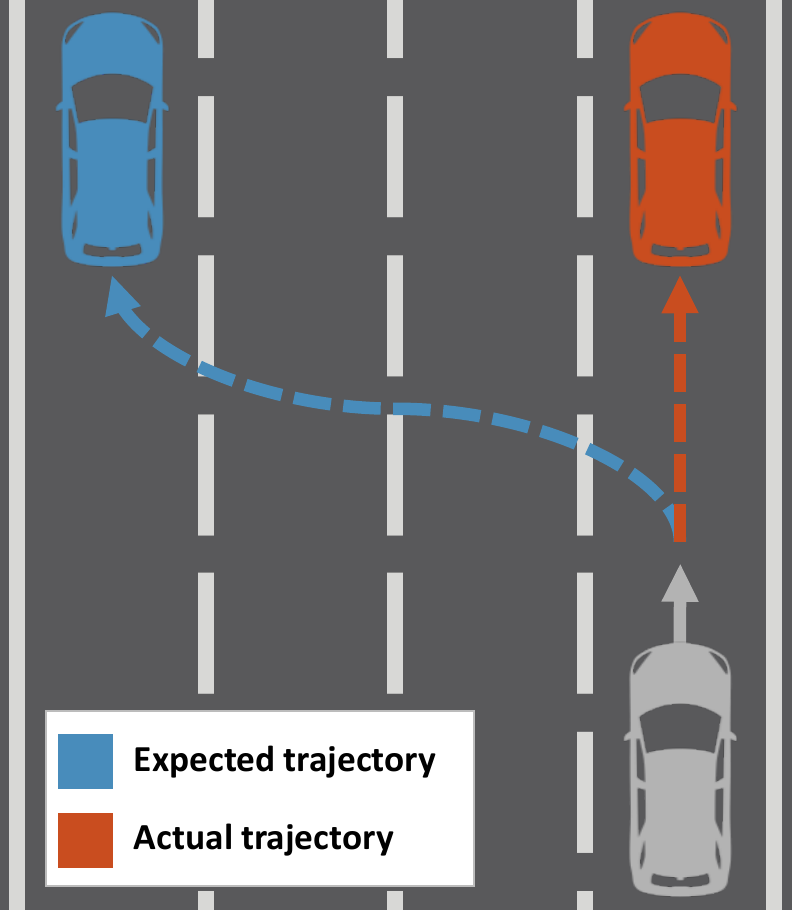}
        \caption{Bug \#6}
        \label{fig:bug6}
    \end{subfigure}
    \begin{subfigure}[b]{0.24\textwidth}
        \centering
        \includegraphics[height=3.3cm]{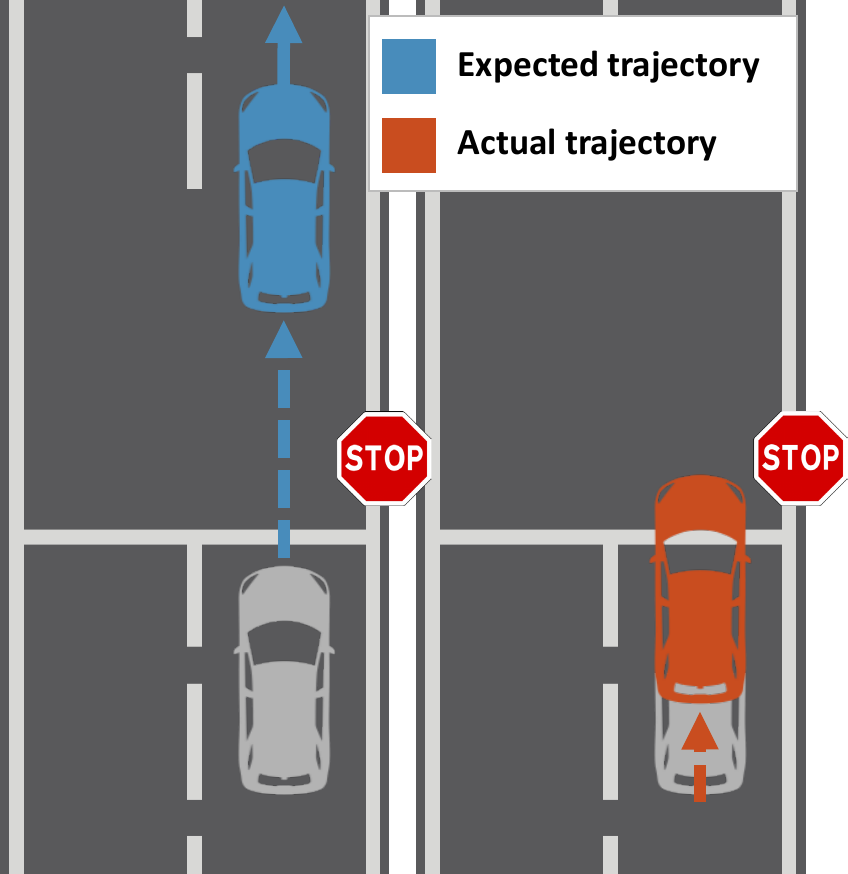}
        \caption{Bug \#5}
        \label{fig:bug5}
    \end{subfigure}
    \begin{subfigure}[b]{0.24\textwidth}
        \centering
        \includegraphics[height=3.3cm]{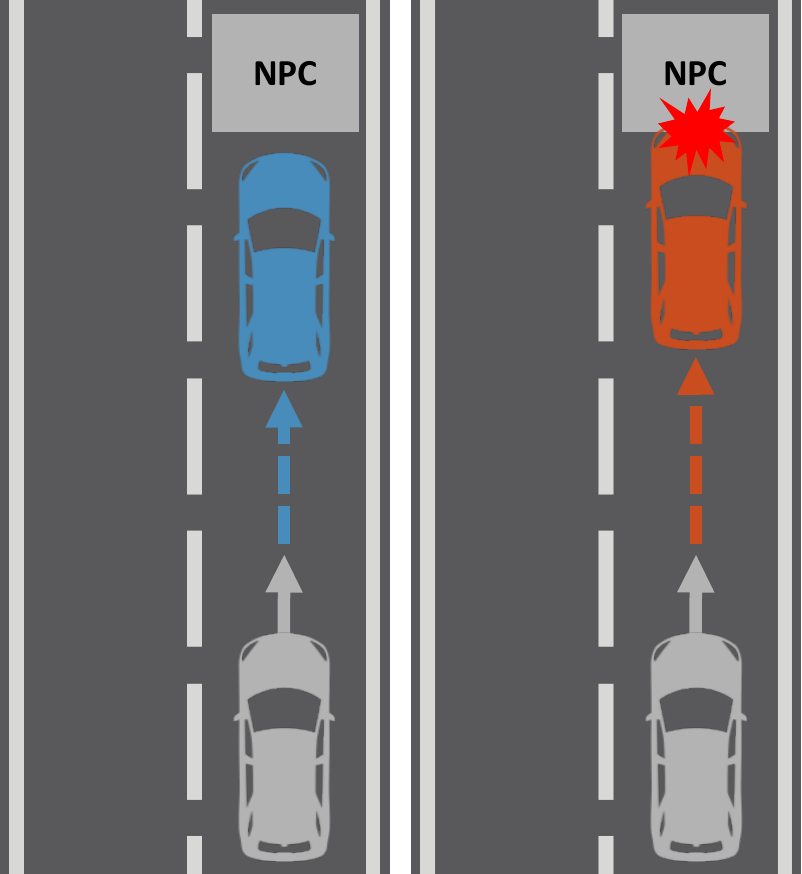}
        \caption{Bug \#4}
        \label{fig:bug4}
    \end{subfigure}
    \begin{subfigure}[b]{0.24\textwidth}
        \centering
        \includegraphics[height=3.3cm]{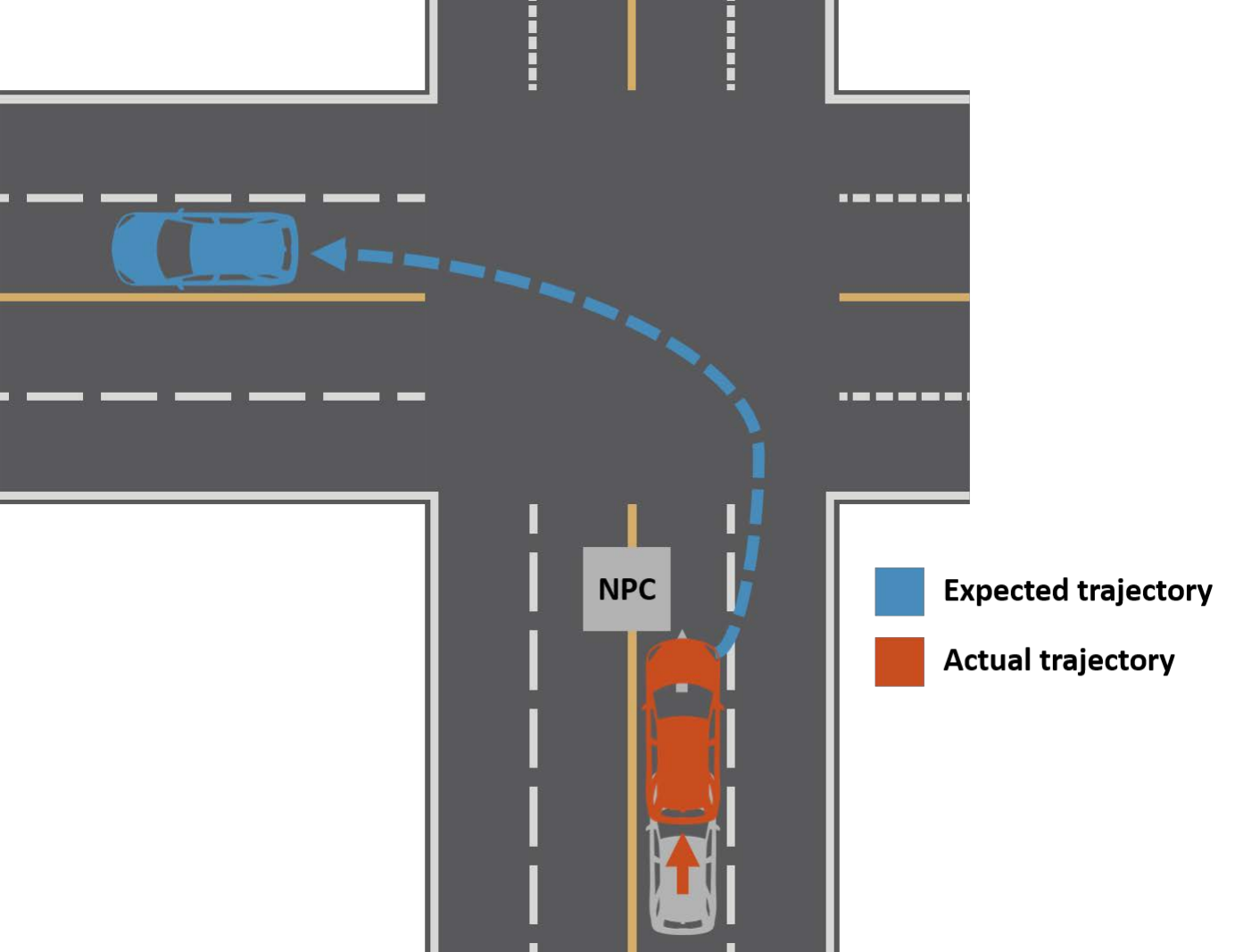}
        \caption{Bug \#8}
        \label{fig:bug8}
    \end{subfigure}
    \caption{Illustration of discovered hazardous behaviors}
    \label{fig:evaluation_case_study}
    \vspace{-4mm}
\end{figure}

\nibf{Case Study 1: Rerouting \& intersection turn failure (Bug \#6).}
Bug \#6 involves a mission failure resulting from the vehicle's inability to reach the correct destination lane as shown in~\cref{fig:bug6}.
In this scenario, the ego vehicle initially starts on the rightmost lane of a multi-lane road but fails to successfully navigate to its intended destination located on the leftmost lane.
This issue originates from a critical bug encountered during the guided path generation process when lane changes are required.

Baidu Apollo's planning system operates through three sequential phases: creating a routing path, generating a guided path, and constructing a trajectory.
When the destination is on a different lane, multiple routes are generated as routing paths.
In Bug \#6, the ego vehicle starts in lane 4 with destination in lane 1, resulting in a routing result of \{lane 4, lane 3, lane 2, lane 1\}.
The reference line generation algorithm determines the feasibility of lane transitions by evaluating two primary criteria through directional and spatial analysis.
In the spatial proximity verification step, it calculates the distance between lane centers using lane width and a predefined buffer zone.
An internal buffer of 0.3 meters is used, with lane width referenced from the HD map.
The maximum allowable distance combines half the width of each lane plus the buffer:

\vspace{-2mm}
\begin{equation}
\vspace{-1mm}
    d < \frac{1}{2}d_{curr, width} + \frac{1}{2}d_{next, width} + d_{buffer}
\end{equation}


In the case of Bug \#6, the HD map defined lane width as 3.5 meters with a 0.3-meter buffer for lane transition calculations.
During operation, the measured distance between lane 4 and lane 3 was 3.88 meters, exceeding the critical threshold of 3.8 meters.
The algorithm's feasibility check prevented reference line generation for lane 3, but critically, the system failed to notify the driver about the inability to perform the requested lane change.
This silent failure represents a significant safety concern, as it could mislead the driver about the vehicle's capabilities and limit situational awareness.
We define this type of error as a configuration error.

\nibf{Case Study 2: Rerouting \& stop sign (Bug \#5).}
Bug \#5 represents a scenario where the ego vehicle fails to resume driving after stopping near a stop sign as shown in~\cref{fig:bug5}.
This issue arises from a rerouting event during an SSU scene, resulting in a scene transition to an LF scene and failure to resume driving.
Two primary factors contribute to this problem: rerouting occurred requiring situation re-evaluation, and an error exists in the decision condition for the SSU scene.

Baidu Apollo uses an indirect stopping mechanism by creating a virtual stop wall in front of the vehicle rather than directly setting speed to zero.
This mechanism applies across obstacle avoidance, traffic lights, and stop signs.
In the case of temporary stopping, such as stop signs, the stop wall is temporarily ignored if the requirements are met, and this process is only performed within the SSU scene.


In this case, the ego vehicle attempts to change lanes and then make a right turn at an intersection with a stop sign.
As the lane change fails, the system initiates rerouting, which clears all planning history and evaluates the current scene.
The scene decision algorithm classifies stop sign scenes based on the vehicle's front end position relative to the stop sign.
In this case, rerouting occurred after the front of the ego vehicle crossed the stop line, so the system triggers LF scene classification instead of SSU scene.
Therefore, Baidu Apollo cannot handle the stop wall and results in permanent stop.

In this scenario, the vehicle encounters a critical conflict: it collides with the stop wall generated by the stop sign, yet cannot transition into the SSU scene necessary for stop wall removal.
This situation creates an unresolvable deadlock.
We categorize such failures, which stem from flaws in the internal conditional logic design, as decision errors.

\nibf{Case Study 3: Front vehicle collision (Bug \#4).}
Bug \#4 demonstrates a scenario where deceleration fails, resulting in a collision with the vehicle ahead as shown in~\cref{fig:bug4}.
Initially, the vehicle ahead is not reliably detected at long distances due to Sense model instability, but becomes reliably detected at approximately 20 meters.
The ego vehicle's maximum deceleration capability is configured at 6~$m/s^2$, which should provide a braking distance of around 6.5 meters—theoretically sufficient to prevent collision.
However, the path optimization process prevented maximum deceleration from being executed, resulting in the collision.
While path optimization prioritizes ride comfort, passenger safety must take precedence.
The ADS should ensure adequate deceleration when abrupt stops are required, prioritizing safety over comfort.
We define this type of error as an optimization error.


\nibf{Case Study 4: Immobile due to front NPC vehicle (Bug \#8).}
Bug \#8 represents a case discovered through the immobile oracle, where the ego vehicle remains stationary near an intersection due to a stopped vehicle ahead as shown in~\cref{fig:bug8}.
Baidu Apollo prohibits all lane change operations, including lane change and lane borrow maneuvers, near intersections.
However, according to official Baidu Apollo documentation, the system is designed to reroute and enable the vehicle to exit the roadway when it remains stationary for an extended period.
The Plan model should provide functionality to handle prolonged stationary situations through mechanisms such as temporarily permitting lane changes to facilitate route exit.
However, the actual implementation lacks such escape functionality for these scenarios.
We define this type of error as an insufficient implementation error, where the system fails to perform intended behavior due to missing functionality rather than incorrect implementation of existing features.



\nibf{Fixed Hazardous Behaviors}
\cref{tab:bug_fix} shows the discovered hazardous behaviors categorized into four error types: 1) Configuration errors involving internal variables and parameter values, 2) Optimization errors in path optimization algorithms, 3) Design errors including incorrect design decisions such as improper branch conditions, and 4) Insufficient implementation representing missing or incomplete functionality.
Optimization errors require algorithmic modifications or parameter adjustments that risk introducing regressions in previously functioning scenarios.
Insufficient implementation issues demand substantial development effort to implement missing functionality.
Configuration and design errors can be addressed through targeted modifications with minimal risk of side effects.

We applied patches to four configuration and design errors, as detailed in~\cref{tab:bug_fix}.
All targeted issues were successfully resolved with no apparent side effects observed during subsequent testing.
For example, Bug \#5 and Bug \#6 both fail to change lane and stop.
After we applied the patch, both scenarios were successfully resolved to perform lane changes at the proper time.
Videos demonstrating system behavior before and after patch application are available at \url{https://sites.google.com/view/safe-planner/home}.

\begin{table}[t]
\scriptsize
\centering

\setlength\tabcolsep{0.030cm}
\def\arraystretch{0.75}

\caption{Error types and fixability of discovered cases}
\label{tab:bug_fix}

\begin{threeparttable}
\begin{tabular}{@{}cllc@{}}
\toprule

Bug ID & Bug Description & Error Type & Fixable? \\


\midrule

01 & Lane borrow immobile/collision & Optimization & \\
02 & Nudge immobile/collision &  Configuration & \checkmark \\
03 & NPC vehicle Cut-in collision & Optimization & \\
04 & Stopped NPC vehicle collision & Optimization & \\
05 & Rerouting\&stop sign & Configuration\&Design & \checkmark \\
06 & Rerouting\&intersection turn failure & Configuration\&Optimization & \checkmark \\
07 & Intersection turn failure & Optimization & \\
08 & Immobile due to front vehicle & Insufficient Implementation & \\
09 & Immobile during stop sign scene & Optimization & \\
10 & Immobile traffic intersection & Optimization & \\
11 & Emergency stop failure & Design & \\
12 & Pull over target change constantly & Configuration & \checkmark \\
13 & Pull over failure due to obstacle & Insufficient Implementation & \\
14 & Pull over failure at intersection & Insufficient Implementation & \\
15 & Destination reach failure & Insufficient Implementation & \\

\bottomrule
\end{tabular}

\vspace{-3mm}
\end{threeparttable}
\end{table}

\begin{answerbox}
\textbf{Answer to RQ1}: \sys detected a total of 520 hazardous behaviors and classified them into 15 cases based on fault localization.
We further fixed 4 cases among the 15 cases, which demonstrates the feasibility of our fault localization approach.
\end{answerbox}
\vspace{-1mm}

\subsection{Code Coverage Analysis}

To evaluate whether \sys covered sufficiently diverse scenarios, we analyzed code coverage metrics.
We implemented instrumentation code to measure both decision coverage and function coverage, tracking accumulated coverage throughout the testing process.
While the Plan model contains numerous helper classes, we focused our coverage analysis on behavior planning components, specifically targeting scene decisions and the modules responsible for scene, stage, and trajectory generation tasks.
We excluded from the coverage analysis several code categories: unreachable code segments, code paths accessible only through configuration changes, and untested scenarios outside our defined scope.
As a result, we found that \sys achieved 83.63\% function coverage and 63.22\% decision coverage.

\begin{figure}[ht]
    \centerline{\includegraphics[width=0.9\columnwidth]{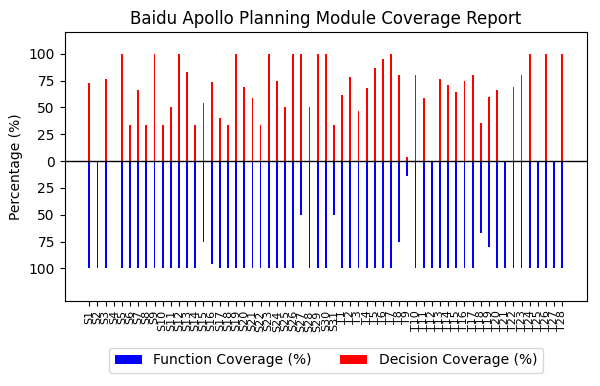}}
    \caption
    {Code coverage result of Baidu Apollo Plan model}
    \label{fig:coverage_result}
    \vspace{-3mm}
\end{figure}

\cref{fig:coverage_result} presents the code coverage analysis results for the target components.
Our coverage analysis yielded several key findings.
First, uncovered decisions were primarily attributed to edge cases including initialization failures and system errors.
Second, we identified a behavioral pattern in the TLURT scene where the creep stage was consistently bypassed.
This scene follows three stages: stop, creep, and intersection\_cruise.
When the ego vehicle's speed exceeds a predetermined threshold during post-stop movement, the system bypasses the creep stage and transitions directly to intersection\_cruise.
Due to configured ego vehicle parameters consistently exceeding this threshold, we observed zero coverage for creep stage-related decisions.
This demonstrates how parameter configuration can prevent execution of specific decision paths.

\begin{answerbox}
\textbf{Answer to RQ2}: \sys provides 83.63\% function coverage and 63.22\% decision coverage.
Uncovered code segments were primarily attributed to edge cases and parameter-dependent paths rather than insufficient scenario diversity, demonstrating the effectiveness of our systematic scenario generation approach.
\end{answerbox}
\vspace{-1mm}

\subsection{Baseline Comparison}
To evaluate the efficiency of \sys, we proposed two design approaches: 1) the systematic generation of driving scenarios, and 2) the scoring metric targeting DDT. 
Given the substantial disparity in the testing scope compared to existing end-to-end ADS testing frameworks, we conducted a baseline comparison to assess the effectiveness of our designs. To facilitate this evaluation, we employed two baseline designs as reference points. This approach allowed us to gauge the performance and capabilities of our designs, considering the unique characteristics of the testing scope.

\nibf{Baselines.}
The initial baseline, $GA\ Only$, replaces the input generation method employed in \sys with random sampling. This substitution allows us to assess the efficacy of our input generation design. The second baseline, $\sys\ without\ GA$, removes the genetic algorithm (GA) component from \sys, enabling us to evaluate the proficiency of our scoring metric within the GA framework. Lastly, the final baseline, $Random$, removes both the input generation and scoring system utilized in \sys. Instead, it relies solely on executing random inputs until hazardous behavior is detected, thereby shedding light on the limitations associated with the random approach.

\nibf{Setup and metrics.}
Both the \sys and baselines were executed for 9 hours each. We utilized two metrics to assess the performance: the number of distinct bugs detected, and the timestamp to identify the most recent bug.

\begin{figure}[t]
     \centering
         \includegraphics[width=0.82\columnwidth]{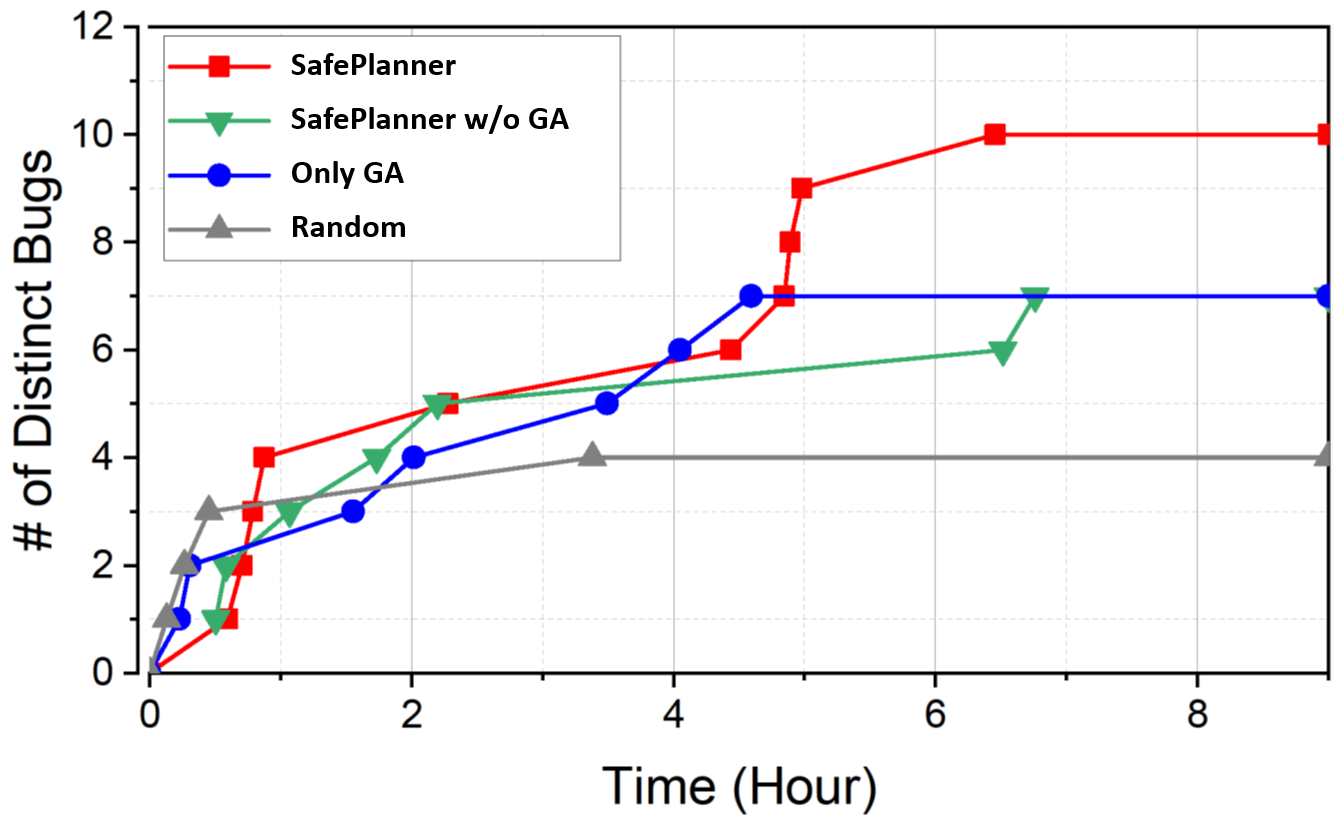}
        \caption{Number of hazardous behaviors detected over the execution of different methods}
        \label{fig:fuzzer_diff}
        \vspace{-5mm}
\end{figure}

\nibf{Results.}
\cref{fig:fuzzer_diff} demonstrates that the $Random$ baseline identifies only 4 bugs, representing 40\% of the bugs detected by \sys.
Additionally, $Random$ ceases to discover new bugs after the 3-hour mark, while \sys continues detecting distinct bugs throughout the 7-hour evaluation period.
These results demonstrate the effectiveness of our approach in discovering hazardous behaviors compared to the baseline.

When considering the baselines incorporating one of our design approaches, a notable degradation in performance is observed compared to \sys.
In terms of the number of distinct bugs detected, both the $Only\ GA$ and $\sys\ w/o\ GA$ baselines exhibit a degradation of 30\%.
However, the $Only\ GA$ baseline reaches its maximum reachability two hours earlier than \sys in this comparison.
Conversely, $\sys\ w/o\ GA$ continues to identify new distinct bugs even after \sys reaches its maximum.
This difference can be attributed to the seed pool, which provides more reproducible scenarios for the baseline methods.

\begin{answerbox}
\textbf{Answer to RQ3}: Both input scenario generation and the fuzzing algorithm are useful for \sys to detect hazardous behaviors.
\end{answerbox}
\section{Related Work}
\label{sec:related}

\subsection{Scenario-based test case generation}
Scenario-based testing in simulation is preferred for ADS evaluation due to the safety risks and high costs of real-world testing.
However, ensuring comprehensive coverage and developing systematic test scenarios remain challenging for effective ADS evaluation.
Previous studies have proposed input generation methods in two directions to tackle this challenge.

The first direction focuses on generating a greater number of safety-critical test scenarios to ensure comprehensive coverage.
Researchers have conducted studies~\cite{tuncali2018simulation, mullins2018adaptive, althoff2018automatic, abeysirigoonawardena2019generating, mullins2017automated, klischat2019generating, ding2020learning, calo2020generating} to increase the quantity and diversity of test scenarios, enabling more thorough evaluation of ADS safety.
The second direction aims to develop more efficient methods for scenario generation.
Researchers have designed algorithms and techniques~\cite{gambi2019automatically, khastgir2017test, tuncali2019rapidly, xia2018test, li2021testing, ponn2019towards, majzik2019towards, wang2021advsim} to generate scenarios more efficiently.
However, previous studies have not practically applied their methodologies to actual business-grade ADS, and they often lack specific focus on the SOTIF standard.

\subsection{ADS Testing Methodologies}
Previous studies have developed methods for different models of the ADS, including sense~\cite{cheng2020towards, dreossi2017systematic, dokhanchi2018evaluating, ramanagopal2018failing, rao2019approach, tu2021exploring, zhou2019automated}, plan \cite{tang2021route, wan2022too, song2023discovering, cao2022advdo, zhang2022adversarial, tan2023targeted}, and end-to-end software \cite{li2020av, kim2022drivefuzz, sun2022lawbreaker, zhou2023specification}. These studies enhance our understanding of ADS performance by validating robustness under various conditions.

For instance, Li \etal focus on detecting collisions caused by changes in nearby vehicles' routes~\cite{li2020av}, while Kim \etal identify potential bugs in the ADS related to road conditions and terrain variations~\cite{kim2022drivefuzz}.
These studies primarily assess ADS robustness in challenging external environments.
In contrast, Wan \etal investigate the Plan model's possible DoS vulnerability conditions by locating objects in non-hazardous ways, emphasizing the significance of behavior planning for ADS performance and safety.
Their approach considers ADS robustness in benign external environments.
\section{Conclusion}
\label{sec:conclusion}
This paper introduces \sys, a comprehensive testing framework for evaluating safety in ADS.
Our framework specifically targets level 4 ADS Plan Model, addressing the fundamental challenges of systematic scenario generation and hazardous behavior detection.
Through in-depth software analysis of Baidu Apollo, we developed a systematic approach for generating input driving scenarios.
Through a guided fuzzing approach, \sys demonstrated its effectiveness by identifying potential issues in the Plan model.
Our evaluation revealed 520 hazardous behaviors across 20,635 test cases, which were systematically classified into 15 distinct categories, 14 of these identified issues were previously unknown.
We conclude by encouraging further research into white-box testing methodologies for autonomous driving systems, as systematic validation of ADS functionality is crucial for advancing toward fully automated vehicles.

\bibliographystyle{IEEEtran}
\bibliography{refs/reference, refs/conf}

\end{document}